\documentclass[10pt,twocolumn,letterpaper]{article}

\usepackage{cvpr}
\usepackage{wrapfig}

\definecolor{agreen}{RGB}{74, 198, 148}
\definecolor{purple}{RGB}{158, 62, 177}
\definecolor{darkpurple}{RGB}{170, 70, 210}
\definecolor{aqua}{RGB}{87, 180, 181}
\definecolor{lightblue}{RGB}{72, 123, 232}
\definecolor{hotpink}{RGB}{255, 83, 115}
\definecolor{teal}{RGB}{90, 200, 250}
\definecolor{linkColor}{RGB}{0, 128, 229}
\definecolor{lightgreen}{RGB}{33, 222, 128}
\definecolor{almostBlack}{RGB}{60,60,60}

\definecolor{red}{RGB}{236, 107, 44}
\definecolor{green}{RGB}{0, 128, 0}
\definecolor{yellow}{RGB}{255, 192, 0}
\definecolor{purple}{RGB}{128, 0, 128}
\definecolor{cyan}{RGB}{0, 255, 255}
\definecolor{lightgray}{gray}{0.95}
\definecolor{grayborder}{gray}{0.5}
\definecolor{gray}{gray}{0.75}
\definecolor{orange}{RGB}{236, 107, 44}
\definecolor{lightorange}{RGB}{255, 223, 186}
\definecolor{blue}{RGB}{116, 95, 232}
\definecolor{lightblue}{RGB}{255, 223, 186}
\definecolor{lightpurple}{RGB}{202,58,126}
\definecolor{benign_purple}{RGB}{150,150,150}
\definecolor{adv_orange}{RGB}{212,64,57}
\definecolor{9colorq1}{RGB}{166,206,227}
\definecolor{9colorq2}{RGB}{31,120,180}
\definecolor{9colorq3}{RGB}{178,223,138}
\definecolor{9colorq4}{RGB}{51,160,44}
\definecolor{9colorq5}{RGB}{251,154,153}
\definecolor{9colorq6}{RGB}{227,26,28}
\definecolor{9colorq7}{RGB}{253,191,111}
\definecolor{9colorq8}{RGB}{255,127,0}
\definecolor{9colorq9}{RGB}{202,178,214}

\newcommand{\github}{\url{https://github.com/poloclub/3D-Gaussian-Splat-Attack}}

\definecolor{codebg}{gray}{0.96}
\definecolor{codeframe}{gray}{0.8}
\definecolor{code_gray}{RGB}{115, 115, 115}
\definecolor{atck_sfc_blue}{RGB}{116, 95, 232}
\definecolor{manipulate_3d_rep_pink}{RGB}{202, 58, 126}
\definecolor{designate_domain_yellow}{RGB}{234, 182, 86}

\newtoggle{inheader}

\definecolor{cvprblue}{rgb}{0.21,0.49,0.74}
\usepackage[pagebackref,breaklinks,colorlinks,allcolors=cvprblue]{hyperref}

\def\paperID{3} 
\def\confName{CVPR}
\def\confYear{2025}

\title{3D Gaussian Splat Vulnerabilities}

\author{Matthew Hull$^1$, Haoyang Yang$^1$, Pratham Mehta$^1$, Mansi Phute$^1$, Aeree Cho$^1$, \\
Haoran Wang$^1$, Matthew Lau$^1$, Wenke Lee$^1$, Willian T. Lunardi$^2$, Martin Andreoni$^2$, Polo Chau$^1$\\
$^{1}$Georgia Tech, $^{2}$Technology Innovation Institute\\
$^{1}${\tt\small [matthewhull, hyang440, pratham, mphute6, aeree, haoran.wang, mattlaued01,} \\
{\tt\small wenke, polo]@gatech.edu},
$^{2}${\tt\small [willian.lunardi, martin.andreoni]@tii.ae}
}

\begin{document}
\twocolumn[{%
\renewcommand\twocolumn[1][]{#1}%
\maketitle
\centering
\vspace{-2.5em}
\includegraphics[width=.85\linewidth]{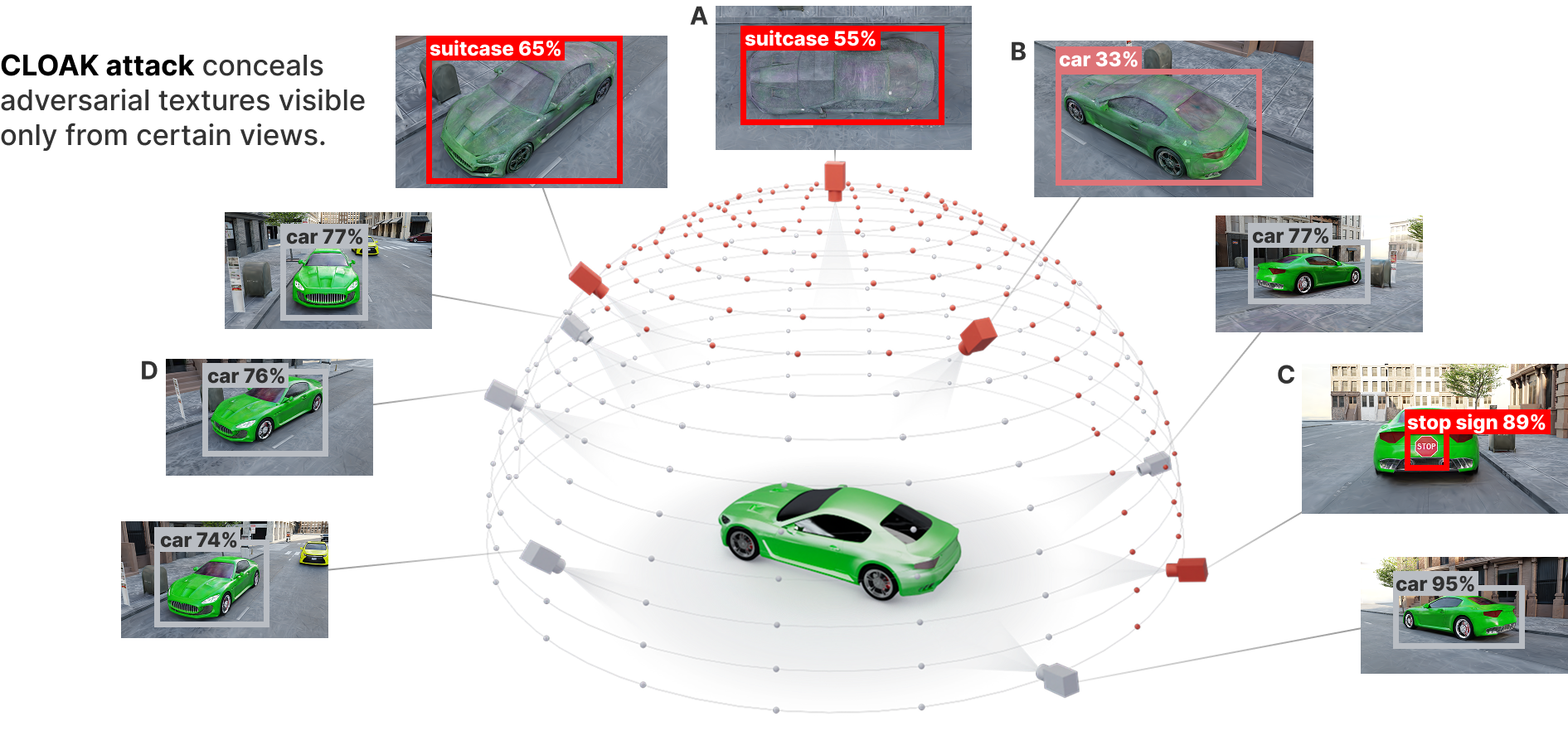}
\vspace{-1.5em}
\captionof{figure}{Our \textbf{CLOAK} attack conceals \textit{multiple} adversarial cloaked textures in 3DGS scenes using Spherical Harmonics, causing the 3DGS representation of the car to become adversarial at different view points (red dots).  For example, (A) when viewed from the top, the car appears as a suitcase, (B) ``car'' detection confidence decreases, (C) and when viewed directly from behind, displays a ``stop sign.''
}
\label{fig:teaser}
\vspace{0.2cm}
}]
\begin{abstract}
With 3D Gaussian Splatting (3DGS) being increasingly used in safety-critical applications, how can an adversary manipulate the scene to cause harm?
We introduce \textbf{CLOAK}, the first attack that leverages view-dependent Gaussian appearances—colors and textures that change with viewing angle—to embed adversarial content visible only from specific viewpoints. 
We further demonstrate \textbf{DAGGER}, a targeted adversarial attack directly perturbing 3D Gaussians without access to underlying training data, deceiving multi-stage object detectors \textit{e.g.}, Faster R-CNN, through established methods such as projected gradient descent.
These attacks highlight underexplored vulnerabilities in 3DGS, introducing a new potential threat to robotic learning for autonomous navigation and other safety-critical 3DGS applications.
\end{abstract}
\vspace{-0.2cm}
\vspace{-0.4cm}
\section{Introduction}
\label{sec:introduction}
3D Gaussian Splatting (3DGS) has rapidly gained popularity due to its efficiency in novel-view synthesis and real-time rendering of complex scenes, outperforming traditional methods like Neural Radiance Fields (NeRFs) \cite{kerbl_3d_2023}. 
These advantages have led to growing interest in safety-critical domains such as autonomous driving \cite{zhou_drivinggaussian_2024,li_vdg_2024}, robotic navigation, and grasping \cite{zheng_gaussiangrasper_2024}, where rapid data generation and accurate sim2real transfer are essential.
A typical 3DGS scene consists of 3D Gaussians initialized from structure-from-motion point clouds, optimized through backpropagation to refine positions, rotations, colors via Spherical Harmonics, scaling, and alpha blending. 
Despite the increasing adoption of 3DGS, vulnerabilities in its optimization processes and representations remain underexplored.
\begin{figure}[t]
    \centering
    \includegraphics[width=\columnwidth]{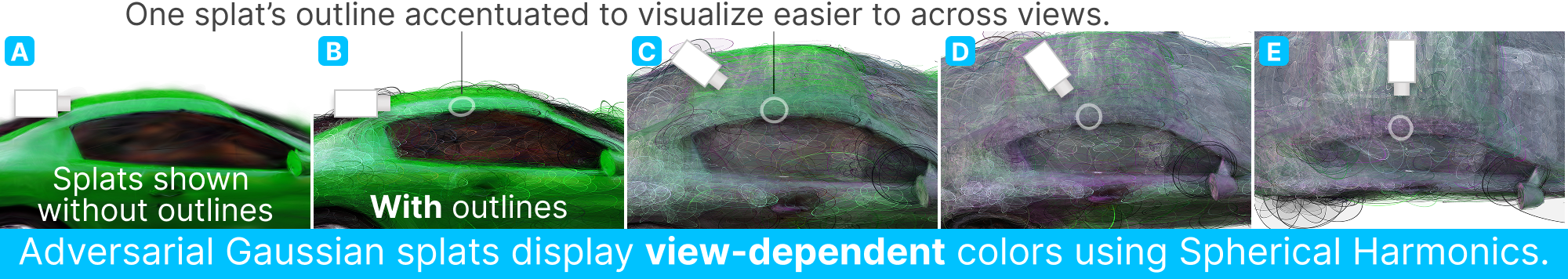}
    \caption{Adversarial Gaussian splats demonstrating view-dependent color changes enabled by spherical harmonic rendering. We highlight a single splat with a light border for easier tracking of color changes across views, revealing its transition from green to gray when rotating from a side view (frames A–B) to an overhead view (frames C–E).}
    \label{fig:fig_2}
    \vspace{-0.5cm}
\end{figure}
We discovered that the view-dependent nature of Spherical Harmonics (SH)—commonly used in real-time rendering for realistic shading, enables adversaries to embed concealed adversarial appearances into 3DGS, each visible only from specific viewing angles (Fig.~\ref{fig:teaser}). 
For instance, an object such as a car could appear benign from ground level yet take on the appearance of asphalt or roadway when viewed aerially, effectively hiding from overhead surveillance systems (see Fig.~\ref{fig:teaser},~\ref{fig:fig_2}). 
Furthermore, gradient-based adversarial methods like Projected Gradient Descent (PGD) can also be generalized to manipulate the Gaussian scene representation directly (Fig.~\ref{fig:fig_3}), causing misclassifications and misdetections in downstream object detection tasks.
Our findings reveal critical yet underexplored vulnerabilities inherent in 3DGS, highlighting a novel avenue for adversarial machine learning research and motivating the need for robust defensive strategies.
To highlight these vulnerabilities, our main contributions are:
\begin{enumerate}
\item \textbf{We introduce the CLOAK attack}--to the best of our knowledge, the first attack to conceal multiple adversarial cloaked textures in 3DGS using Spherical Harmonics, causing the scene to become adversarial at different view points. 
We demonstrate CLOAK on YOLOv8, causing missed detections and misclassifications.
CLOAK stands for \textit{\textbf{C}oncealed \textbf{L}ocalized \textbf{O}bject \textbf{A}ttack \textbf{K}inematics}.
\item \textbf{We introduce the DAGGER attack, a generalization of the PGD technique to 3DGS scenes}. 
DAGGER directly manipulates 3DGS, targeting two-stage object detection models such as Faster R-CNN without needing access to the original image training data.
DAGGER stands for \textit{\textbf{D}irect \textbf{A}ttack on \textbf{G}aussian \textbf{G}radient \textbf{E}vasive \textbf{R}epresentations}. 
\item \textbf{An open-source implementation on GitHub\footnote{\github{}}} to support reproducibility, further research, and defense development. 
\end{enumerate}
\section{Related Work}
\label{sec:related_work}
\vspace{-0.2cm}
Adversarial attacks in the 2D space are well-established, and the corresponding vulnerabilities are extensively studied. 
However, such studies are not prevalent regarding 3D spaces \cite{li_survey_2024}.
Recently, differentiable renderers have been used to perform gradient optimization of components in a scene, which can be used to create highly realistic scenes where perturbations are applied to geometry, texture, pose, lighting, and sensors. 
This results in physically plausible objects that could be transferred to the real world.
Adversarial ML researchers have also recently investigated exploiting novel views in NeRFs to create template inversion attacks to fool facial recognition systems \cite{shahreza_comprehensive_2023}.
e.g., synthesizing novel views from limited data, and gaining access to systems using a 3D model of a face and the resulting new views.
Importantly, these attacks do not require white-box access to the targeted model weights, highlighting a lower barrier for adversaries and raising concerns due to their practical feasibility.
To date, only two works have explored limited threat model vulnerabilities in 3DGS. 
One introduces a computational cost attack targeting the split/densify stages of the 3DGS algorithm by perturbing training images, significantly increasing training time, scene complexity (in terms of Gaussian count), and memory usage, while reducing rendering frame rates; however, this approach does not target downstream models or tasks \cite{lu_poison-splat_2024}. 
The second work targets only a single model (CLIP ViT-B/16), employing data poisoning through segmentation and perturbation of target regions within images to induce targeted and untargeted misclassifications, and it does not directly manipulate the underlying 3DGS scene representation \cite{zeybey_gaussian_2024}.
\section{Attack Methods}
\label{sec:attack_methods}
\subsection{Threat Models}
\vspace{-0.2cm}
\label{subsec:threat_models}
3DGS synthesizes novel views by training a volumetric representation (using Gaussians and SH coefficients) from images, presenting adversaries with vulnerabilities at different pipeline stages (Fig~\ref{fig:teaser}). 
Our \textbf{CLOAK} attack models an adversary who can only manipulate training data, embedding concealed adversarial content visible solely from specific viewpoints, without direct access to internal scene parameters.

\begin{figure}[t]
    \centering
    \includegraphics[width=\columnwidth]{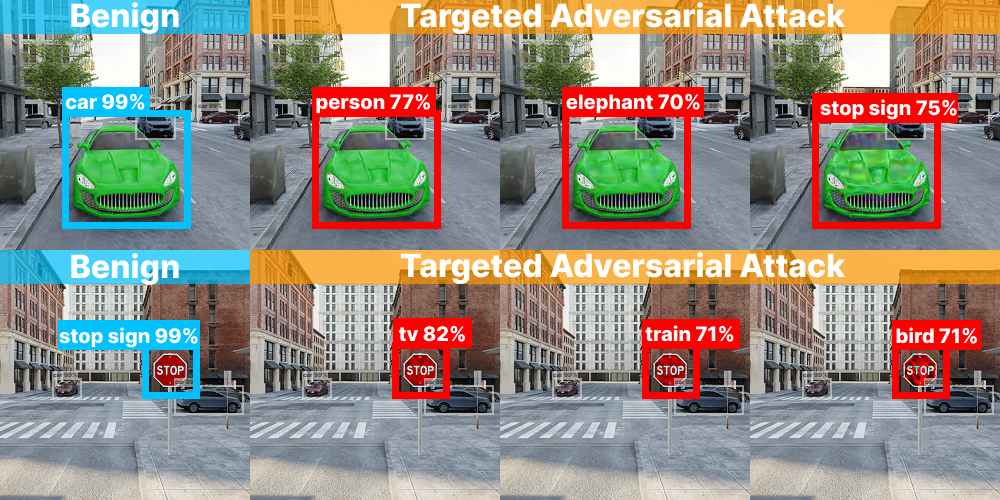}
    \caption{\textbf{DAGGER} manipulates Gaussian attributes to induce misdetections on Faster R-CNN. On the top row, the car's color is perturbed in a targeted attack, resulting in high-confidence misclassifications as a ``person'', ``elephant'', and ``stop sign.''. In the second row, the stop sign is attacked, causing the model to misclassify it as a ``tv'', ``train'', and ``bird''.}
    \label{fig:fig_3}
    \vspace{-0.5cm}
\end{figure}

In contrast, the \textbf{DAGGER} attack considers a stronger adversary who directly modifies the Gaussian representation, optimizing parameters like position, SH, scaling, rotation, and transparency. The resulting manipulated scene is rendered and passed to a downstream object detection model, causing targeted or untargeted misclassifications (Fig.~\ref{fig:fig_3}).
\subsection{CLOAK Attack}
\label{subsec:cloak_method}
\vspace{-0.20cm}
Our CLOAK attack leverages the view-dependent appearance properties of 3DGS to conceal adversarial content within seemingly benign 3D scenes. 
By exploiting SH encoding, we can create objects with different appearances based on viewing angle.

In 3DGS, each Gaussian is assigned SH coefficients rather than a fixed RGB color. 
These SH functions define how color varies with the incident viewing direction, allowing a Gaussian's appearance to change dynamically depending on the observer's perspective. 
During training, SH encode color information for varying camera views, enabling scenes to appear benign or adversarial depending on viewpoint.

To hide adversarial views within an object, we begin with a benign textured version of a 3D model alongside one or more adversarial textures. 
A training image dataset is created by rendering the object with benign textures from one set of camera views and adversarial textures from targeted camera views. 
The attack trains the 3DGS scene so that certain viewpoints appear completely normal while others reveal hidden adversarial content.

This technique enables sophisticated concealment. 
For example, a car can be designed with an adversarial appearance from a top view while maintaining benign appearances from all other angles (Fig.~\ref{fig:teaser},Fig.~\ref{fig:fig_4}). 
Walking 360 degrees around such a vehicle on the ground appears completely normal, as the top of the car viewed from ground level shows no indication of the hidden adversarial content.

We formulate our CLOAK attack as follows. Let $\mathcal{D} = \{(x_i, c_i)\}_{i=1}^N$ be the benign dataset, where each image $x_i \in X$ is associated with a camera pose $c_i \in C$. 
The attacker selects a subset of targeted camera poses $C^* \subset C$ and generates adversarial images $\tilde{x}_i$ for each viewpoint $c_i \in C^*$, modifying the appearance of a target object while preserving the scene’s visual realism. 
The attack replaces each original image $x_i$ with its adversarial counterpart $\tilde{x}_i$ for $c_i \in C^*$, forming the attacked dataset $\mathcal{D}' = \{(A(x_i, c_i), c_i)\}_{i=1}^N$, where  
\vspace{-0.15cm}
\begin{equation}
    \vspace{-0.15cm}
    A(x, c) = 
    \begin{cases} 
        \tilde{x}, & \text{if } c \in C^*, \\
        x, & \text{otherwise}.
    \end{cases}
\end{equation}

\noindent Training the 3DGS model on $\mathcal{D}'$ ensures that from non-targeted viewpoints $c \notin C^*$, the target object retains its benign appearance, while from viewpoints $c \in C^*$, the adversarial modifications become embedded in the learned scene. 
This results in an attack that remains concealed under initial observations but reveals manipulated content from attacker-specified angles.

\subsection{DAGGER Attack}
\label{subsec:dagger_method}
\begin{figure}[t]
    \centering
    \includegraphics[width=\columnwidth]{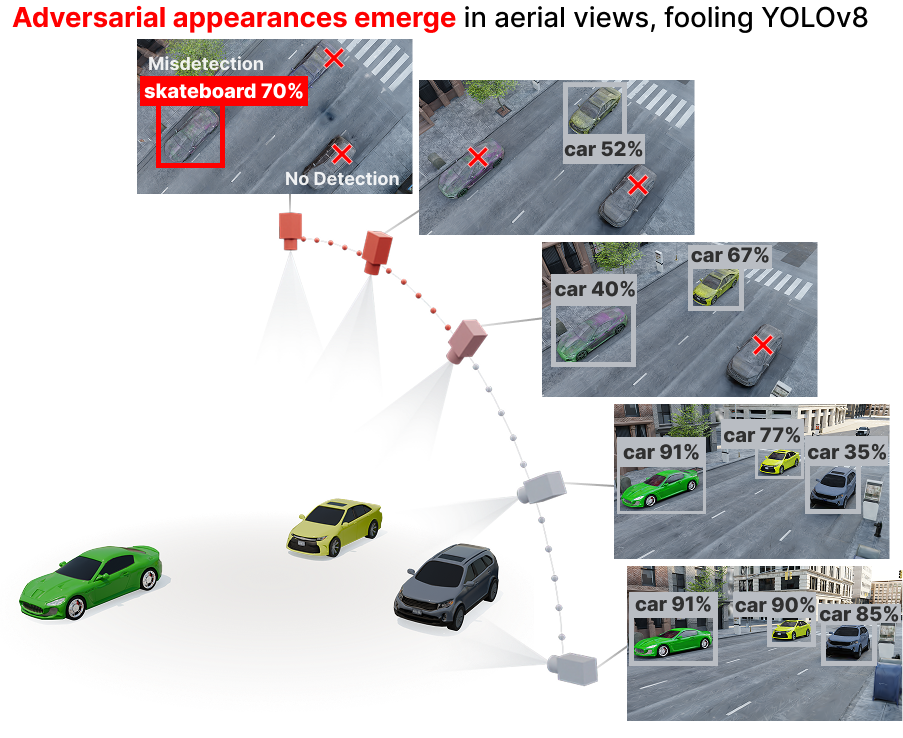}
    \caption{YOLOv8 detections over adversarial viewpoints attacked by \textcolor{adv_orange}{CLOAK}.}
    \label{fig:fig_4}
    \vspace{-0.6cm}
\end{figure}

The DAGGER attack assumes a more powerful adversary with access to the 3DGS scene representation and a target downstream model (Fig.~\ref{fig:fig_4}).

Unlike CLOAK, this attack does not require access to training data, assuming white-box access to the scene and downstream model.
A 3DGS scene is comprised of a data structure holding attributes of each 3D Gaussian to represent: 
SH coefficients (color) $\mathbf{c}$, 
$xyz$ coordinates $\mathbf{p}\in\mathbb{R}^3$, 
scaling factor $s$, 
rotation $r$, and 
transparency $\alpha$. 
Training of a 3DGS scene uses differentiable rendering, meaning that gradients flow to Gaussian attributes to iteratively adjusting them to represent the training data in a process similar to backpropagation for training a deep neural network. 
Borrowing from existing adversarial gradient optimization attacks on 2D images \cite{madry_towards_2018} (and 3D scenes), we know that an attacker with access to a target model, can optimize the scene representation (already shown in differentiable rendering attacks). 
Suppose this attacker can access the 3DGS scene file. In that case, they can carry out a gradient optimization PGD attack by targeting one or more 3DGS attributes and optimizing it to maximize some loss function.

In our DAGGER attack, let $\mathcal{G} = \{g_1,\ldots,g_n\}$ be the set of 3D Gaussians, where each $g_i = (\mathbf{p}_i,\mathbf{c}_i,s_i,r_i,\alpha_i)$.
A differentiable renderer $R(\mathcal{G})$ maps these parameters to 2D images, which are then passed to a downstream model $M$. 
The adversary selects a subset $\Theta$ of parameters to manipulate, aiming to maximize a loss $\mathcal{L}\bigl(M(R(\mathcal{G})),y\bigr)$ under a constraint $\|\Theta - \Theta_0\|\le \epsilon$. Formally,
\vspace{-0.15cm}
\begin{equation}
    \vspace{-0.15cm}
    \max_{\mathcal{G}'}\,\ell\!\bigl(M\bigl(R(\mathcal{G}')\bigr), y\bigr)\quad\text{subject to}\quad\|\Theta-\Theta_0\|\le\epsilon,
\end{equation}
and uses a projected gradient step
\vspace{-0.15cm}
\begin{equation}
\vspace{-0.15cm}
    \Theta_{t+1} \leftarrow \Pi_{\|\Theta-\Theta_0\|\le\epsilon} \Bigl(\Theta_t + \eta\,\nabla_{\Theta_t}\,\ell\bigl(M(R(\mathcal{G})), y\bigr)\Bigr),
\end{equation}
where $\Theta_0$ are the original parameters, $\eta$ is the step size, and $\Pi$ is the projection operator. This iterative procedure yields a modified $\mathcal{G}'$ whose rendered output misleads $M$.

\section{Experiments}
\label{sec:Experiments}

\subsection{CLOAK Experiments}
We conducted experiments using Blender (\small{\url{www.blender.org}}) \normalsize with the Cycles renderer to create photorealistic renderings of a car captured from 210 distinct camera angles covering a hemispherical region, enabling complete 360-degree visualization (Fig.~\ref{fig:teaser}).
We embedded three concealed adversarial appearances among benign views: a normal appearance from 110 angles (Fig.~\ref{fig:teaser}D), a ``road'' texture at 80 overhead angles (Fig.~\ref{fig:teaser}A), and a ``stop sign'' texture at 20 angles directly behind the car (Fig.~\ref{fig:teaser}C).
Training the 3DGS scene with this dataset successfully created an object whose concealed adversarial textures emerged distinctly from specific viewpoints—overhead for the ``road'' texture and rear angles for the ``stop sign.''
Diagonal views obscure these adversarial modifications, potentially misleading both human observers and object detector into assuming consistency across viewpoints.

To evaluate attack effectiveness, we conducted a black-box assessment using YOLOv8 object detection.
The scene was rendered with camera viewpoints smoothly transitioning from benign ground-level angles toward adversarial overhead and rear angles.
Rendered frames analyzed by YOLOv8 (Fig.~\ref{fig:fig_4}) demonstrated significant reductions in detection confidence, including complete missed detections when adversarial textures were fully visible.
In particular, YOLOv8 detected the car successfully from 80 out of 110 benign viewpoints but failed to detect it in 78 out of 80 adversarial overhead (``road'') views.

\subsection{DAGGER Experiments}
In our direct attack experiments targeting 3D Gaussians (Fig.~\ref{fig:fig_3}), we began by rendering a 3D scene in two parts, creating a composite scene.
We maintained a Gaussian splat index corresponding to the targeted object splats while masking gradients for all non-targeted scene splats, ensuring that perturbations and optimizations were applied exclusively to the targeted object.
For each targeted viewpoint, we perturbed the color attributes of the Gaussians using SH coefficients, controlling the perceived RGB color from specific angles.
Using white-box access to a Faster R-CNN object detection model, we iteratively rendered the composite scene, computed the detection loss, and applied projected gradient descent (PGD) updates to the SH coefficients.
After each perturbation step, the adjusted SH coefficients are converted to RGB during rasterization, and the scene is re-rendered for subsequent optimization steps.
This method effectively enabled targeted manipulation of object appearance from specified viewpoints, significantly influencing object detection outcomes.
For example, we successfully optimized Faster R-CNN to misclassify a ``car'' as an ``person'' with consistently high detection confidence ($>70\%$) in just 11 iterations using PGD $\ell_2$-norm, with attacker budget $\epsilon=5.0$, and learning rate $\alpha=\epsilon\cdot2/\mathrm{steps}$. 
\section{Conclusion and Ongoing Work}
\label{sec:conclusion}
In this paper, we demonstrated unexplored vulnerabilities in the emerging 3D Gaussian Splatting (3DGS) framework, highlighting security implications for safety-critical applications. 
Our proposed CLOAK and DAGGER attacks show how adversaries can exploit training-time and post-training vulnerabilities to deceive state-of-the-art object detection models. 
We release our methods openly to support future research on securing 3DGS-based systems.
{
    \small
    \bibliographystyle{ieeenat_fullname}
    \bibliography{main}
}

\end{document}